\documentclass{aa}
\bibpunct{(}{)}{;}{a}{}{,}

\usepackage{graphicx}
\usepackage{txfonts}
\usepackage{hyperref}
\usepackage[normalem]{ulem}

\begin{document}

\title{Detectability of stochastic gravitational wave background from weakly hyperbolic encounters}

\author{Morteza Kerachian \inst{\ref{ASU}} \thanks{E-mail: kerachian.morteza@gmail.com} \and
Sajal Mukherjee \inst{\ref{ASU},\ref{Birla}} \and Georgios Lukes-Gerakopoulos\inst{\ref{ASU}}  \and Sanjit Mitra\inst{\ref{IUCAA}}}
\institute{Astronomical Institute of the Czech Academy of Sciences, Bo\v{c}n\'{i} II 1401/1a, CZ-141 00 Prague, Czech Republic \label{ASU}
\and
Birla Institute of Technology and Science Pilani, Rajasthan, 333031, India \label{Birla} 
\and
Inter-University Centre for Astronomy and Astrophysics (IUCAA) Pune, Ganeshkhind, Maharashtra 411007, India \label{IUCAA}
}

\authorrunning{M. Kerachian et al.}
\titlerunning{Detectability of the SGWB from weakly hyperbolic encounters}

%\email{kerachian.morteza@gmail.com}

% Abstract of the paper
 \abstract
{ 
We compute the stochastic gravitational wave (GW) background generated by black hole--black hole (BH--BH) hyperbolic encounters with eccentricities close to one and compare them with the respective sensitivity curves of planned GW detectors. We use the Keplerian potential to model the orbits of the encounters and the quadrupole formula to compute the emitted GWs. We take into account hyperbolic encounters that take place in clusters up to redshift $5$ and with BH masses spanning from $5 M_{\odot}$ to $55 M_{\odot}$. We assume the clusters to be virialized and study several cluster models with different mass and virial velocity, and finally obtain an accumulative result,  displaying the background as an {average}. Using the maxima and minima of our accumulative result for each frequency, we provide analytical expressions for both optimistic and pessimistic scenarios. Our results suggest that the background from these encounters is likely to be detected by the third-generation detectors \textit{Cosmic explorer} and \textit{Einstein telescope}, while the tail section at lower frequencies intersects with \textit{DECIGO}, making it a potential target source for both ground- and space-based future GW detectors.
}

\keywords{black hole physics -- gravitational waves -- scattering -- globular clusters: general}

 \maketitle

\nolinenumbers

%%%%%%%%%%%%%%%%%%%%%%%%%%%%%%%%%%%%%%%%
\section{Introduction}
%%%%%%%%%%%%%%%%

Up to this point, terrestrial gravitational wave (GW) observatories have detected only binary mergers of stellar compact objects \citep{LIGOVIRGO_O1_O2,LIGOVIRGO_O3a,LIGOVIRGO_O3b}. The signal from these sources remains discrete, in the sense that it does not overlap with the signal from another source when detected. This is expected to change in the future \citep{Maggiore:2020,LISA}, when GW observatories are expected to receive signals from multiple sources simultaneously. A more extreme case of such overlap is seen when the sources are so numerous that one cannot be discerned from another, creating a stochastic GW background (SGWB).

There are several types of SGWBs depending on their sources \citep{StochLISA,StochTer,NANOGrav15SMBH,NANOGrav15NP,ePTA}; they can be split into two categories: cosmological sources and astrophysical sources. The latter category contains mainly different types of compact binary coalescences, but also hyperbolic encounters of compact objects. Contrary to the compact binary coalescences for which the signal can last for many cycles, in the case of hyperbolic encounters the signal can be viewed rather as a transient \citep{Morras:2022}. The peak of this transient GW signal lies near the pericenter of the respective hyperbolic trajectory \citep{DeVittori:2012}, which should correspond to short bursts of GWs. The exact waveform of these bursts \citep{Cho:2018,Vines:2019,Jakobsen:2021,Saketh:2022,Chowdhuri:2023} is not relevant for the study of an SGWB, because we are interested in the accumulative effect of many events.

It is often argued that bound compact clusters, such as globular clusters (GCs), can be one of the possible dominant channels for hyperbolic encounters in the Universe \citep{Dymnikova:1982,Kocsis:2006}. Therefore, encounters in GCs are expected to provide an adequate number of events for an SGWB \citep{Mukherjee:2021}. This rate depends on the number of black holes (BHs) in a cluster and on the number of clusters in a galaxy. Therefore, from an astrophysical point of view, detecting a SGWB or failing to do so would put constraints on the numbers mentioned above. 
 
Our work continues the investigation done by \citet{Mukherjee:2021}, in which an estimate of the detectable event rates was provided. In particular, we investigate BH hyperbolic encounters where eccentricities tend to be almost parabolic; that is, the eccentricity is very close to one. We obtain different energy density profiles of the SGWB generated by these encounters while varying the parameters of the system, such as the masses of the BHs and the virial velocity in the cluster. The comparison of these profiles with the respective sensitivity curves of GW observatories allows us to infer some preferable scenarios for detection.

The remainder of the article is organized as follows: in Sect.~\ref{sec:Anal} we describe how we modeled the orbital dynamics of the hyperbolic encounters and the respective GW radiation. In  Sect.~\ref{sec:StochBack} we show how we computed the dimensionless GW energy density spectrum for the hyperbolic encounters. In Sect.~\ref{sec:Results} we present our results and in Sect.~\ref{sec:Disc} we conclude with a discussion of our main findings.

%%%%%%%%%%%%%%%%%%%%%%%%%%%%%%%%%%%%%%%%%%
\section{Analytical calculations} \label{sec:Anal}
%%%%%%%%%%%%%%%%%%
\subsection{Orbital parameters}\label{sec:orbitalparameter}
%%%%%%%%%%%%%%%%%%

\begin{figure}
    \begin{center}
        \includegraphics[width=0.35\textwidth]{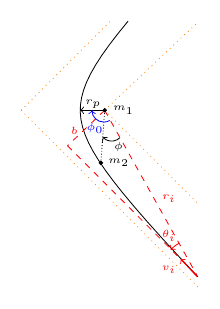}
    \caption{Hyperbolic trajectory of a secondary compact object of mass $m_2$ in the gravitational field of a BH of mass $m_1$. The secondary object is initially located at a distance $r_i$ from the primary;  the initial angle between the line connecting the bodies and the initial velocity $v_i$ is $\theta_i$. The motion of the secondary can be parametrized by the radial distance $r(\phi)$ as a function of the azimuthal angle $\phi$, where  $\phi$  lies between $r_i$ and $r(\phi)$. The $r_p$ is the periapsis, which is the minimum distance between the bodies,  and  $\phi_0$ is the periapsis angle, which is the angle between $r(\phi)$ and $r_p$. As shown here, we define the impact parameter $b$ as the vertical distance between the primary BH and the initial velocity of the secondary;  that is, $b=r_i \sin \theta_i$.}\label{fig:trajectory}
    \end{center}
\end{figure}

In this section, we briefly review the orbital parameters for hyperbolic encounters limited within a radius of a cluster, as given by \citet{Mukherjee:2021}. This motion is planar and each orbit is determined by its initial conditions: the initial distance $r_i$ between the bodies, the initial angle $\theta_i$ between $r_i$  and the initial velocity of the secondary $v_i$, and the masses of the objects $m_1$ and $m_2$ (for an illustration see Fig.~\ref{fig:trajectory}). As mentioned above, the initial radius is chosen to be less than the cluster radius, that is $r_i\leq R_c$, where $R_c$ is the radius of a cluster; $v_i$ is estimated from virial theorem; and  $\theta_i$ is considered to have an arbitrary value within its physical range \footnote{This range is discussed in section~\ref{sec:SpeCas}.}. 

As the objects are far from each other, we use the Keplerian potential energy. The Hamiltonian for this system is 
\begin{equation}\label{eq:hamiltonian}
    H= E= \frac{1}{2}( p_r^2+\frac{L^2}{r^2})- \frac{G M}{r}, 
\end{equation}
where $E$, $p_r$, and $L$ are the total energy, the radial momentum, and the total angular momentum per unit of reduced mass $\mu$, while $M=m_1+m_2$ is the total mass and $1/\mu=1/m_1+1/m_2$ is the reduced mass of the system. Moreover, from Fig.~\ref{fig:trajectory} the total angular momentum per $\mu$ can be written as $L= r_i v_i \sin \theta_i$. Solving Hamiltonian~\eqref{eq:hamiltonian} allows us to find that 
\begin{equation}\label{eq:radtraj}
    r(\phi)=\frac{p}{1+e \cos(\phi-\phi_0)},
\end{equation}
where $\phi$ is the angle between $r_i$ and $r(\phi)$ and $p$, $e,$ and $\phi_0$ are given by
\begin{equation}\label{eq:p&e&tanphi}
 p= \frac{b^2}{a},\quad e^2=1+\dfrac{b^2}{a^2}\alpha, \quad \tan \phi_0=-\dfrac{b}{a}\beta,
\end{equation}
%%%%
with 
\begin{align}\label{eq:transformation}
    a&=GM/v_i^2, \qquad \quad b= L/v_i,\nonumber\\
    \alpha&=1-2a/r_i,\qquad \beta=\cos\theta_i (1-b^2/(a r_i))^{-1}.
 \end{align}
See~\citet{Mukherjee:2021} for a detailed derivation of these quantities\footnote{In Appendix~\ref{sec:hypermotion}, we show that this trajectory is a part of a hyperbola with different parametrization, namely with semi-major axis $a/\alpha$, and semi-minor axis $b/\sqrt{\alpha}$.}.

%%%%

\subsection{Energy radiation in the time domain}
%%%%%%%%%%%%%%%

As we are dealing with an SGWB, to calculate the gravitation radiation reaction from the hyperbolic encounters, it is reasonable to assume that an estimation of the leading terms of this reaction is sufficient \citep{Roskill:2023}. Therefore, we use the quadrupole formula \citep[see, e.g.,][]{Maggiore:2007ulw} should be sufficient for the scope of the present work.  The quadrupole moment tensor is given by $$Q_{i j }= \mu ( 3 x_i x_j- \mid r\mid^2 \delta_{i j}),$$ where the nonzero components of this system are 
%%%%%%
\begin{align}
    Q_{11}&=\mu r^2 (3 \cos^2\phi-1),\label{eq:q11t}\\
    Q_{22}&=\mu r^2 (3 \sin^2\phi-1),\label{eq:q22t} \\
  Q_{12}&= Q_{21}=3 \mu r^2 \cos\phi \sin\phi,\label{eq:q12t}\\
  Q_{33}&=-\mu r^2.\label{eq:q33t}
\end{align}
%%%%%%
Then, the power emitted through  GW is determined from
%%%%%%%%%
\begin{align}
    P=&\dfrac{dE}{dt}=-\dfrac{G}{45 c^5}(\dddot{Q}_{ij}\dddot{Q}^{ij})=-\dfrac{4L^6G}{15 p^8 \mu^4 c^5}[1+e \cos(\phi-\phi_0)]^4\nonumber\\
    &\big\{1+13 e^2+ 48 e\cos(\phi-\phi_0)+11e^2\cos(2\phi-2\phi_0)\big\}.
\end{align}
%%%%%%%%%
By substituting the relations~\eqref{eq:transformation}, we arrive at
%%%%%%%%
\begin{equation}\label{eq:powertime}
    P(t)= \dfrac{dE}{dt}=-\dfrac{32G}{45 c^5}\dfrac{\mu^2 \alpha^4 v_i^6}{b^2}f(\phi,e)
,\end{equation}
%%%%%%%%
with
%%%%
\begin{align}
    f(\phi,e)&=\frac{3 \left(1+e \cos(\phi-\phi_0)\right)^4}{8 (e^2-1)^4}\Big\{24+13 e^2\nonumber\\
    &+48 e \cos(\phi-\phi_0)+11 e^2 \cos2(\phi-\phi_0)\Big\}.
\end{align}
%%%%
The above expression exactly matches the existing literature, except for $\alpha$, which arises from the different definitions of $a$ and $b$ (see Appendix~\ref{sec:hypermotion})~\citep{juan}.

We note that the relation between $\phi$ and $t$ can be derived from $L= r^2 \dot{\phi}$. To get an explicit relation, one can rewrite this in the form 
%%%%
\begin{equation}
    L t= p^2 \int^{\phi}_{\phi 0} \dfrac{d\phi}{[1+e\cos(\phi-\phi_0)]^2},
\end{equation}
%%%%
and consequently, we get 
%%%%%%
\begin{align}\label{eq:phit}
    Lt&=- p^2 \Big[\dfrac{2}{(e^2-1)^{3/2}}\tanh^{-1} \Big(\sqrt{\dfrac{e-1}{e+1}}\tan\big(\dfrac{\phi-\phi_0}{2}\big)\Big)\Big]\nonumber\\
    +&\dfrac{ p^2 e \sin(\phi-\phi_0)}{(e^2-1)(1+e \cos(\phi-\phi_0))}.
\end{align}

%%%%%%

\subsection{Energy radiation in the frequency domain}\label{sec:FT}

To compute the frequency spectrum of the radiated power  $\Delta E$, we employ the Fourier transformation (FT) of the energy emission in the time domain; that is,
\begin{equation}\label{eq:totalE}
    \Delta E= \int^{+\infty}_{-\infty} P(t) dt= \frac{1}{\pi}\int^{+\infty}_{0} P(\omega) d\omega,
\end{equation}
where $P(\omega)$ is the power radiation in the frequency domain given by
\begin{equation}\label{eq:powerspectrum}
    P(\omega)=\frac{G}{45 c^5} \sum_{i,j}  \widehat{\dddot{Q_{i j}}}\widehat{\dddot{Q_{i j}}}^{*},
\end{equation}
and the over-hat represents the FT defined as
\begin{equation}\label{eq:defft}
    \widehat{f(\omega)}= \int^{+\infty}_{-\infty} f(t) \boldsymbol{e}^{- i \omega t} dt.
\end{equation}
Additionally,  Eq.~\eqref{eq:defft} implies $\widehat{\dot{f}}= i \omega \widehat{f}$, and consequently $\widehat{\dddot{f}}= -i \omega^3 \widehat{f}$. For simplicity, we use this relation for Eq.~\eqref{eq:powerspectrum} and write it in terms of $Q_{i j}$ as follows:
\begin{equation}\label{eq:powerft}
    P(\omega)= \frac{G \omega^6}{45 c^5}  \widehat{Q_{i j}}\widehat{Q_{i j}}^{*}.
\end{equation}
The radial trajectory given in Eq.~\eqref{eq:radtraj} is a function of $\phi$ and the dependence of $\phi$ in terms of $t$ is given by Eq.~\eqref{eq:phit}. However, this parameterization is not suitable if the FT is going to be applied. Therefore, we reparameterize the hyperbolic trajectory in terms of the \textit{eccentric anomaly} $\xi$ as follows:
\begin{equation}\label{eq:trajxi}
    r=a_{\rm c}(e \cosh\xi-1),\qquad \omega_0 t=(e \sinh\xi-\xi),
\end{equation}
where $a_c=a/\alpha,$ and 
\begin{equation}
    \omega_0=\sqrt{\dfrac{GM}{a^3_{\rm c}}}
\end{equation}
is the Keplerian frequency. 

In Cartesian coordinates,  the trajectory~\eqref{eq:trajxi} reads
\begin{equation}\label{eq:cartesianxi}
    x=a_{\rm c}\big(e-\cosh\xi\big), \quad y=a_{\rm c}\sqrt{e^2-1}\sinh\xi.
\end{equation}
Expressing the trajectory in Cartesian coordinates allows us to rewrite the quadrupole momentum tensor~\eqref{eq:q11t}-~\eqref{eq:q33t}  in terms of $\xi$. Namely, we have that
\begin{align}
    Q_{11}&= \frac{1}{2} a_c^2 \mu \left(1+5 e^2- 8 e \cosh \xi+ (3-e^2) \cosh 2\xi \right),\label{eq:q11}\\
    Q_{22}&=  \frac{1}{2} a_c^2 \mu \left(1-4 e^2+ 4 e \cosh \xi- (3-2 e^2) \cosh 2\xi \right),\label{eq:q22}\\
    Q_{12}&=Q_{21}= \frac{1}{2} a_c^2 \mu\left(3 \sqrt{e^2-1} \left( 2 e \sinh \xi- \sinh 2 \xi \right) \right),\label{eq:q12}\\
    Q_{33}&= \frac{1}{2} a_c^2 \mu \left(-2 -e^2+ 4 e \cosh\xi - e^2 \cosh 2\xi\right).\label{eq:q33}
\end{align}

By using FT, whose derivation can be found in Appendix~\ref{sec:ftderivation}, and the fact that $ \nu= \omega/\omega_0,$ the power radiation~\eqref{eq:powerft} becomes\footnote{We ignore the constant terms in the quadrupole momentum tensor because they make no contribution to the total energy; i.e., $\displaystyle \int_0^\infty \omega^3 \widehat{f(c)} d\omega=0$, where $\widehat{f(c)}$ is the FT of the constant terms.  An FT of a constant term contains a delta Dirac of $\omega$; i.e.,  $\displaystyle \int_0^\infty\omega^3 \delta(\omega) d\omega=0.$}
\begin{equation}\label{eq:powerf}
    P(\omega)= \frac{16 \pi^2}{180}\frac{ G^3 M^2 \mu^2}{ a_c^2\, c^5} \nu^4 \boldsymbol{f}(e,\nu),
\end{equation}
where
\begin{align}\label{eq:fenu}
    &\boldsymbol{f}(e,\nu)=\Big| \frac{3(e^2-1)}{e} H^{(1)\prime}_{i \nu}(i \nu e)+ \frac{e^2-3}{e^2} \frac{i}{\nu}H^{(1)}_{i \nu}(i \nu e)\Big|^2\nonumber\\
    +&\Big|\frac{3(e^2-1)}{e} H^{(1)\prime}_{i \nu}(i \nu e)+ \frac{2 e^2-3}{e^2} \frac{i}{\nu}H^{(1)}_{i \nu}(i \nu e)\Big|^2+\frac{18 (e^2-1)}{e^2}\nonumber\\
    \times& \Big| \frac{e^2-1}{e} H^{(1)}_{i \nu}(i \nu e)- \frac{i}{\nu} H^{(1)\prime}_{i \nu}(i \nu e) \Big|^2+\Big|\frac{i}{\nu} H^{(1)}_{i \nu}(i \nu e) \Big|^2.
\end{align}
Equation~\eqref{eq:fenu} can be approximated by~\citep{juan}
\begin{equation}\label{eq:fenuapp}
    F(e,\nu)=\nu^4 \boldsymbol{f}(e,\nu)= \frac{12 F(\nu)}{\pi y (1+y^2)^2} e^{-2 \xi(y) \nu},
\end{equation}
where
\begin{align}
    F(\nu)&= \nu ( 1-y^2-3 \nu y^3+4 y^4+9 \nu y^5+6 \nu^2 y^6),\\ \nonumber
    \xi(y)&= y-\arctan y,\\ \nonumber
    y&= \sqrt{e^2-1}. \nonumber
\end{align}
Consequently, the total energy in the frequency domain~\eqref{eq:totalE} can be obtained from Eq.~\eqref{eq:fenuapp} and Eq.~\eqref{eq:powerf}  and is
\begin{align}\label{eq:TEFapp}
    \Delta E&= \frac{1}{\pi} \int_0^\infty P(\omega) d\omega = \frac{16 \pi}{180}\frac{ G^{7/2} M^{5/2} \mu^2}{ a_c^{7/2}\, c^5} \int_0^\infty F(e,\nu) d\nu.
\end{align}

%%%%%%%%%%%%%%
\section{Stochastic gravitational wave background from hyperbolic encounters} \label{sec:StochBack}
%%%%%%%%%%%%%%%
To evaluate the SGWB from hyperbolic encounters, we compute the dimensionless GW energy density spectrum as ~\citep{juan2022,Maggiorebookv2}
%%%%%%%%%%%
\begin{equation} \label{eq:Omegaf}
    \Omega_{\rm GW}(f)=\dfrac{1}{\rho_c}\int^{\infty}_{0}\dfrac{dz}{1+z} \int d\xi \frac{d N(z;\xi)}{d\xi}\dfrac{dE_{\rm GW}(f_r;\xi)}{d \ln f_r}\Big|_{f_r},
\end{equation}
%%%%%%%%%%%
 where $\xi=\lbrace \xi_{1},\, .\,.\,.\,, \xi_m\rbrace$ is a collection of different encounters and
\begin{equation}
    d\xi \frac{d N(z;\xi)}{d\xi} =d\xi_1\, .\,.\,.\,d\xi_m \frac{d N(z;\xi_1,\, .\,.\,.\,, \xi_m)}{d\xi_1\, .\,.\,.\, d\xi_m},
\end{equation}
where $f_r= f (1+z)$ is the GW frequency at the source frame, $dE_{GW}/d \ln f_r$ is the GW energy emission per logarithmic frequency bin in the source frame, $\rho_c$ is the energy density of the Universe given by $\rho_c=3H^2_0 c^2/8\pi G$, $H_0=100 h_{100}~\text{km/sec/Mpc}$ is the Hubble constant, and $N(z;\xi)$ is the number of GW events density at redshift $z$, as given by~\citep{Zhu:2012xw}:
\begin{equation}\label{eq:neventrate}
    N(z;\xi)  =\dfrac{\mathcal{R}(z; \xi)}{(1+z)H(z)},
\end{equation}
in which  $\mathcal{R}(z; \xi)$ is the event rate density and $H(z)=H_0 \sqrt{\Omega_{\Lambda}+ \Omega_m (1+z)^3}$ is the Hubble parameter at $z$. By assuming a standard $\Lambda$ cold dark matter ($\Lambda$CDM) cosmology  we have $\Omega_\Lambda=0.685$, and $\Omega_m=0.315$.

We define the event rate density $\mathcal{R}(z; \xi)$ as 
\begin{equation}\label{eq:eventratedens}
    \mathcal{R}(z; \xi)=\dfrac{P_{\rm clus}(z; \xi) n_{\rm gc}\mathcal{N}(z)}{(1+z) V},
\end{equation}
where $P_{\rm clus}(z; \xi)$ is the probability per unit time of the encounters inside each GC, $n_{\rm gc}$ is the number of GCs per Milky Way-equivalent galaxy (MWEG), $\mathcal{N}(z)$ is the number of MWEGs between redshift $z$ and $z+ dz$, $V$ is the comoving volume up to redshift $z$, and the factor $(1+z)$  represents the cosmological time dilation.  Therefore, the right-hand side of the Eq.~\eqref{eq:eventratedens} provides us with the total event rate density. 

By substituting Eq.~\eqref{eq:eventratedens} and Eq.~\eqref{eq:neventrate} into Eq.~\eqref{eq:Omegaf}, we arrive at
\begin{align}\label{eq:omgwf2}
     \Omega_{\rm GW}(f)=\dfrac{n_{\rm gc}}{\rho_c}&\int^{\infty}_{0}\dfrac{\mathcal{N}(z) dz}{(1+z)^3 V H(z)} \times \nonumber \\
     &\int d\xi \frac{d P_{\rm clus}(z;\xi)}{d\xi}\dfrac{dE_{\rm GW}(f_r;\xi)}{d \nu} \nu.
\end{align}
In Eq.~\eqref{eq:omgwf2}, we take advantage of the fact that
\begin{equation}
    \dfrac{dE_{\rm GW}(f_r;\xi)}{d \ln f_r}= f_r \dfrac{dE_{\rm GW}(f_r;\xi)}{d f_r}= \nu \dfrac{dE_{\rm GW}(\nu;\xi)}{d \nu},
\end{equation}
where $\nu= 2 \pi \nu_0 f_r= 2 \pi \nu_0 f (1+z)$~\footnote{This relation is obtained from the fact that $\omega= 2 \pi f_r$, in which we substitute $\omega=\nu \omega_0$ and use the relation $\nu_0=1/\omega_0$. We note that $\nu$ is an auxiliary variable and not a physical variable.}, $\nu_0=1/\omega_0$, and $dE_{\rm GW}(\nu;\xi)/d\nu$ is given by Eq.~\eqref{eq:TEFapp}.
%%%%%

The second integral in Eq.~\eqref{eq:omgwf2} is of particular interest, and we can call it $\mathcal{K}(z)$ for future reference, that is,
\begin{equation}\label{eq:defkz}
    \mathcal{K}(z)=\int d\xi \frac{d P_{\rm clus}(z;\xi)}{d \xi}\frac{dE_{\rm GW}(\nu;\xi)}{d\nu} \nu.
\end{equation}
We need to integrate the above expression over $\xi$, which in our case is related to $r_i$ and $\theta_i$, that is, $\xi=\xi(r_i,\theta_i)$, to add the contribution from all the radial and angular distributions. 

 To derive the $P_{\rm clus}(z;r_i, \theta_i)$, let us assume that there is an interaction inside a cluster, and the relative initial angle is $\theta_i,$ as shown in Fig. \ref{fig:trajectory}. The probability of such encounters is then 
\begin{equation}
    P_{\theta}=\dfrac{1}{4\pi}\int 2 \pi \sin\theta_i d\theta_i
.\end{equation}

However, this encounter can take place at a given (relative) distance, $r_i$, from the primary body. Therefore, we may want to integrate in a radial direction to add all the contributions in a given volume:
\begin{align}
    P_{\rm indv}&=\int_{r_i} \int_{\theta_i}\dfrac{1}{4\pi}(2\pi \sin\theta_i)d\theta_i (4\pi r_i^2 n_s)dr_i,\nonumber\\
    &=\int_{r_i} \int_{\theta_i}2\pi n_s r_i^2 \sin\theta_id\theta_i dr_i.
\end{align}
We note that in the above expression, $n_s$ is the number density of compact objects inside the cluster. For the number of compact objects, $n_{\rm co}$, we have $n_s=3 n_{\rm co}/(4 \pi R^3_{\rm co})$, where $R_{\rm co}$ is the GC core radius. It is reasonable to express the probability of an encounter per unit of time, that is as a rate of encounters. For this, we impose the relation 
\begin{align}\label{eq:EncR}
    \bar{P}_{\rm indv} 
    = \int_{r_i} \int_{\theta_i} \frac{2\pi n_s}{t_{\rm col}} r_i^2 \sin\theta_id\theta_i dr_i ,
\end{align}
where $t_{\rm col}$ is the collision time. Now, if we ignore the relativistic corrections, then $t_{\rm col} \leq r_i/v_i$, where $v_i$ is the initial velocity, which in our case is the virial velocity. We note that the maximum value of $t_{\rm col}$ is $r_i/v_i$, and this therefore expresses the lowest possible rate. This means that we end up with a conservative rate estimate. 

By employing our assumption in Eq.~\eqref{eq:EncR}, we can now obtain the total rate of the entire cluster, as $P_{\rm indv}$ is the probability per object. By ignoring the small-scale structure, we have
%%%
\begin{align}\label{eq:pclus}
    P_{\rm clus}(z;r_i, \theta_i)=& n_{\rm co} \bar{P}_{\rm indv} \nonumber \\
    =& \dfrac{3 n^2_{\rm co}(z)}{2 R^3_{\rm co}} \int_{r_i}\int_{\theta_i} r_i   \sin\theta_i v_i d\theta_i dr_i.
\end{align}
Therefore, by substituting Eq.~\eqref{eq:pclus} into Eq.~\eqref{eq:defkz}, we arrive at
\begin{align}\label{eq:fkzthet0}
\mathcal{K}(z)=& \dfrac{3 n^2_{\rm co}(z)}{2 R^3_{\rm co}} \times\nonumber\\
&\int_{r_i}\int_{\theta_i}\Big( r_i\sin\theta_iv_i\Big)\Big(\frac{dE_{\rm GW}(\nu_i;r_i,\theta_i)}{d\nu_i}\Big)_{r_i,\theta_i} \nu_i d\theta_i dr_i,
\end{align}
where  $\nu_i$ is a function of $r_i$, because $\nu_i= 2 \pi \nu_{{0_i}} f (1+z)$ with $\nu_{0_i}= \sqrt{a_{c_i}^3/G M}$ and $$a_{c_i}=\frac{a_i}{\alpha_i}=\frac{G M r_i}{r_i v_i^2-2 G M}.$$

At this point, we rewrite Eq.~\eqref{eq:fkzthet0} in terms of the orbital parameter, that is, eccentricity, instead of the angular coordinate $\theta_i$. From~\eqref{eq:p&e&tanphi} and~\eqref{eq:transformation}, we have 
%%%
\begin{equation}\label{eq:thethae0}
    \sin \theta_i=\frac{G M}{v_i} \sqrt{\frac{e_i^2-1 }{r_i (r_i v_i^2-2 G M)}}.
\end{equation}
Substituting $\theta_i= \Theta(e_i, r_i, v_i)$ and $d\theta_i= \Hat{\Theta}(e_i, r_i, v_i) de_i$ from  Eq.~\eqref{eq:thethae0}, into Eq.~\eqref{eq:fkzthet0}, and  we get
%%%%
\begin{align}\label{eq:fkzer0}
 &\mathcal{K}(z)=  \dfrac{3 G^2 M^2 n^2_{\rm co}(z)}{2 R^3_{\rm co}} \int_{r_i} \int_{e_i} dr_i de_i \frac{dE_{\rm GW}(\nu_i;r_i,e_i)}{d\nu_i} \mathcal{F},
\end{align}
%%%
where
%%%
\begin{equation}
    \mathcal{F}=\frac{(e_i\, r_i^2\, v_i\, \nu_i)\Big(r_i ( r_i v_i^2- 2 G M)\Big)^{-1/2}}{\Big(G^2 M^2(e_i^2-1)-r_i v_i^2 (r_i v_i^2-2 G M)\Big)^{1/2}}.
\end{equation}
%%%

Moreover, $\mathcal{N}(z)$ is determined from
%%%%
\begin{equation}\label{eq:Nz}
    \mathcal{N}(z)=\dfrac{4\pi r^2_z c}{H(z)}n(z),
\end{equation}
%%%%
where  $n(z)$ is the number of MWEGs per volume at redshift z, and we assume it to be constant at $n(z)=0.01 \rm{Mpc}^{-3}$\citep{sedda2020}.

Finally, we derive $\Omega_{\rm GW}(f)$ by substituting Eq.~\eqref{eq:fkzer0} and Eq.~\eqref{eq:Nz} into Eq.~\eqref{eq:omgwf2}, and using $V=\dfrac{4}{3}\pi r^3_{\rm zmax}$ as the comoving volume and $r_{\rm zmax}$ as the comoving distance:
%%%%%%
\begin{equation}\label{eq:fomegagw0}
     \Omega_{\rm GW}(f)=\dfrac{n_{\rm gc}}{V \rho_c} \int^{\infty}_{0}\dfrac{\mathcal{N}(z) \mathcal{K}(z) dz}{(1+z)^3  H(z)}. 
\end{equation}
%%%%%%
In order to obtain $r_{\rm zmax}$, we use the equation
%%%%
\begin{equation}
    \dfrac{dr}{dz}=\dfrac{c}{H(z)},
\end{equation}
%%%%
from which we can see that redshift $z_{\rm max}=5$ corresponds to a distance of $r_{\rm zmax}=7.84~\text{Gpc}$.

    \begin{figure*}
        \centering
        \includegraphics[width=0.45\textwidth]{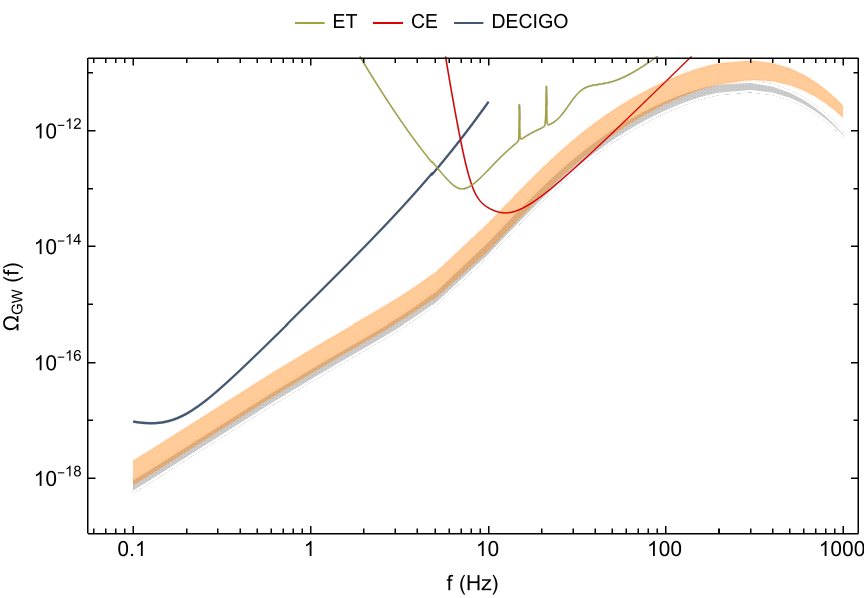}
        \includegraphics[width=0.45\textwidth]{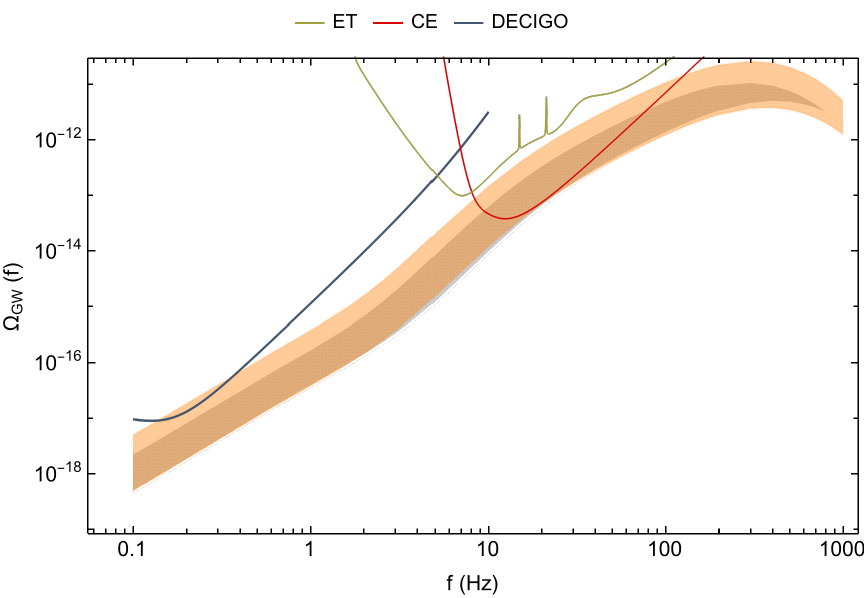}
         \includegraphics[width=0.45\textwidth]{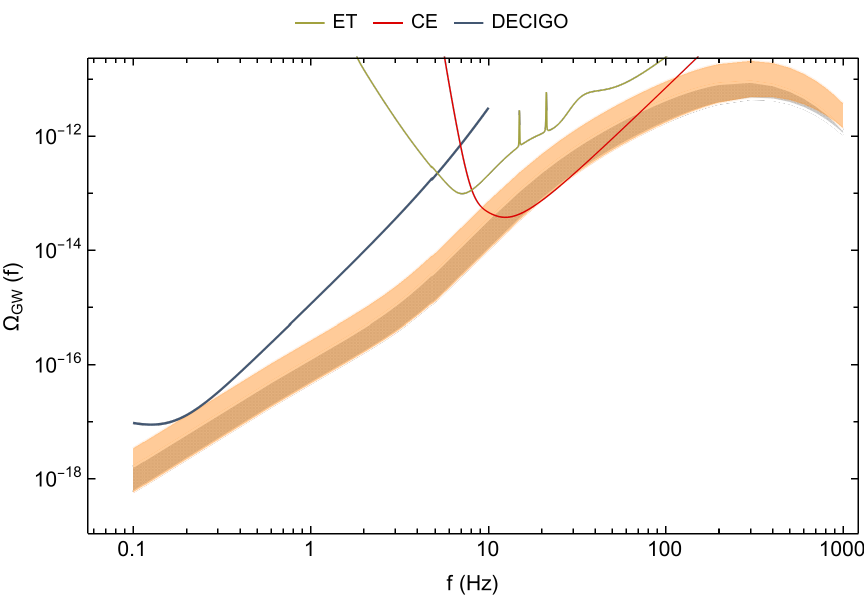}
        \includegraphics[width=0.45\textwidth]{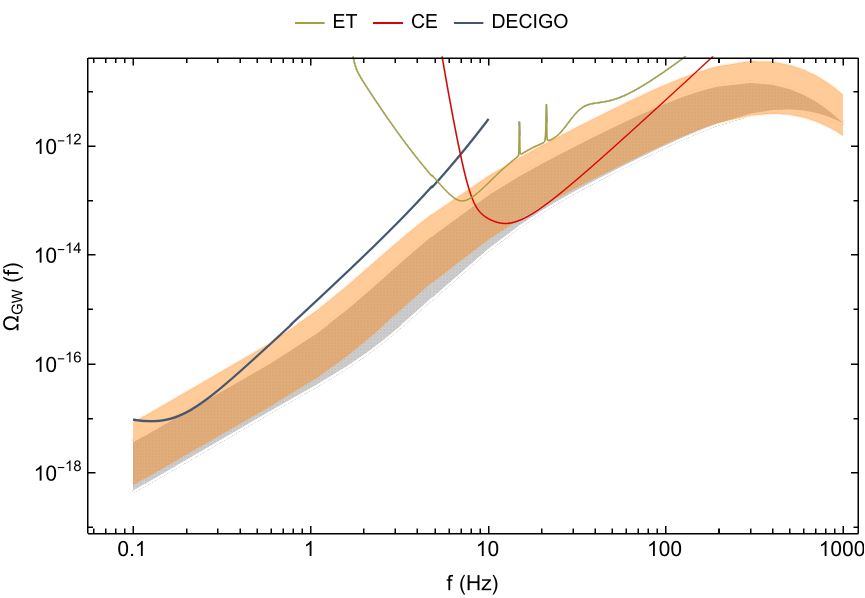}
        \caption{ SGWB as detected using weakly hyperbolic encounters. The shaded regions show the SGWB and are produced from Eq.\eqref{eq:fomegagw0}, where we set $r_i$ in two sets of ranges. The small range varies from $\lbrace 0.03 {\rm pc}, 0.53 {\rm pc}\rbrace$, and the larger range varies from $\lbrace 0.03 {\rm pc}, 9.03 {\rm pc}\rbrace$. In both cases, the eccentricity covers the range of $\lbrace e_{i\rm min}, e_{i\rm min}+ 4\times 10^{-9}\rbrace$. For the top-left plot, we set $N_{\rm tot}=10^6$, $R_{\rm gc}=10 \rm pc$ (see case~\ref{lb:1} in the text for more details.); for the top-right plot, we set $N_{\rm tot}=5\times 10^5$, $R_{\rm gc}=10 \rm pc$ (see case~\ref{lb:2} for more details); for the bottom-left plot, we set $N_{\rm tot}= 10^6$, $R_{\rm gc}=15 \rm pc$ (see case~\ref{lb:3} for more details); for the bottom- right plot, we set $N_{\rm tot}=5\times 10^5$, $R_{\rm gc}=15 \rm pc$ (see case~\ref{lb:4} for more details). In all four cases, we integrate up to redshift $z=5$. The gray region shows the case for the small $r_i$ and the orange region shows the case for the larger $r_i$. Moreover, ET, CE, and DECIGO curves correspond to the sensitivity of the Einstein Telescope \citep{Punturo:2010zza}, the Cosmic Explorer~\citep{cosmicexplorer}, and the Deci-hertz Interferometer Gravitational-wave Observatory \citep{DECOGI} detectors, respectively, for one year of observation~\citep{oneyearsens}. }
            \label{fig:background}
    \end{figure*}

%%%%
\section{Results} \label{sec:Results}
%%%%

In the previous section, we derive the energy density GW spectrum using Eq.~\eqref{eq:fomegagw0} under certain assumptions for a collection of hyperbolic encounters. In this section, we illustrate GW spectra for various parameter options to investigate how certain parameter choices influence these spectra. We also provide an accumulative case representing the reasonable combinations we can take into account to generate an SGWB from hyperbolic encounters.

\subsection{Specific cases} \label{sec:SpeCas}

In order to observe how changing the parameters in our setup varies the detectability of the SGWB, we fix some parameters and vary others, as follows:
\begin{itemize}
\item We set the number of compact objects $n_{\rm co}(z)$ from~\citet{chatterjee2020}.

\item We fix the periapsis $r_{\rm p}$ to be no smaller than the Schwarzschild radius; that is, $r_{\rm p}\geq  r_{\rm sch}$, in which we exclude any head-on collision.

\item We evaluate $\mathcal{K}(z)$ from Eq.\eqref{eq:fkzer0} in a small range $\lbrace r_{i\rm min}, r_{i\rm max} \rbrace= \lbrace 0.03\rm pc, 0.53\rm pc \rbrace$, and a larger range $\lbrace r_{i\rm min}, r_{i\rm max} \rbrace= \lbrace 0.03\rm pc, 9.03\rm pc \rbrace$. In both cases, the eccentricity varies from $\lbrace e_{i \rm min}, e_{i\rm max} \rbrace= \lbrace e_{i \rm min}, e_{i \rm min}+ 4\times 10^{-9} \rbrace$, where $e_{i\rm min}$ comes from $r_{i \rm p}= r_{\rm sch}$; that is, by using relation~\eqref{eq:radtraj} for $\phi=\phi_0$, we get
\begin{align}\label{eq:rp=rsch}
 r_{i\rm p}=\frac{p_i}{1+e_{i\rm min}}= \frac{2 G m_1}{c^2} = r_{\rm sch}.
\end{align}
 From~\eqref{eq:p&e&tanphi} and~\eqref{eq:transformation}, we have $p=b^2/a$, $a= G M/v_i^2$, and $b=L/v_i$; moreover, the total angular momentum can be written as $L= r_i v_i \sin \theta_i$. Therefore, $b$ can be written as $b= r_i \sin \theta_i$ and $p_i= v_i^2 r_i^2 \sin^2 \theta_i/ G M$. By substituting Eq.~\eqref{eq:thethae0}, we get
 \begin{equation}
     r_{i\rm p}= \frac{G M r_i (e_{i}-1)}{r_i v_i^2- 2 G M}.
 \end{equation}
 Therefore, from Eq.~\eqref{eq:rp=rsch} for $e_{i\rm min}$ at each $r_i$, we get
 \begin{equation}
     e_{i\rm min}= 1+ \frac{2 m_1}{M c^2 r_i} (r_i v_i^2- 2 G M).
 \end{equation}

\item  We set the primary mass $m_1$ to be no greater than $55 M_\odot$, and we set the secondary mass $m_2$ to be no greater than $25 M_\odot$ for the small range of $r_i$, and no greater than $20 M_\odot$ for the larger range of $r_i$~\citep{Kocsis:2006}.

 \item We assume that the cluster is virialized to start with, and we have
%%%
\begin{equation}\label{eq:virialv}
    v_i=v_{\rm vir}= \sqrt{\frac{ G N^{\rm tot} \langle m \rangle}{3 R_{\rm gc}}},
\end{equation}
%%%
which is the virial velocity for a uniformly distributed mass~\citep{longair}. In Eq.~\eqref{eq:virialv}, $\langle m \rangle$ is the average mass of a star, and we keep $\langle m \rangle= M_{\odot}$ to be constant throughout our calculation. However, we vary the total number of stars $N^{\rm tot}$, the cluster's radius $R_{\rm gc}$, and $m_1$ and $m_2$  as follows:
\begin{enumerate}
    \item \label{lb:1} \textit{Case $1$: } We choose $N^{\rm tot}= 10^6 $ and $ R_{\rm gc}=10~\rm pc$ and evaluate Eq.~\eqref{eq:fomegagw0}. We observe that in this case for the range of small $r_i$ (gray area in the top-left plot of Fig.~\ref{fig:background}) the SGWB  can barely be detected by Cosmic Explorer (CE) for mass ranges $48 M_\odot \lesssim m_1 \leq 55 M_\odot$ and $19 M_\odot \lesssim m_2 \leq 25 M_\odot$. However, for the range of larger $r_i$ (orange area in the left top plot of Fig.~\ref{fig:background}),  we observe that the  SGWB should be detectable by CE  in the mass ranges $37 M_\odot \lesssim m_1 \leq 55 M_\odot$ and $12 M_\odot \lesssim m_2 \leq 20 M_\odot$.

    \item \label{lb:2} \textit{Case $2$: } We set $N^{\rm tot}= 5 \times 10^5 $ and $ R_{\rm gc}=10~\rm pc$. In this case, the respective SGWBs should be detectable by CE for the range of small $r_i$ (gray area in the top-right plot of Fig.~\ref{fig:background}) and for mass ranges $31 M_\odot \lesssim m_1 \leq 55 M_\odot$ and $12 M_\odot \lesssim m_2 \leq 25 M_\odot,$ and for the range of larger $r_i$ (orange area in the right top plot of Fig.~\ref{fig:background}) and mass ranges $25 M_\odot < m_1 \leq 55 M_\odot$ and $7 M_\odot \lesssim m_2 \leq 20 M_\odot$. SGWBs within the range of large $r_i$ may only be barely detectable by the Deci-hertz Interferometer Gravitational-wave Observatory (DECIGO).
    
    \item \label{lb:3} \textit{Case $3$: } We assume $N^{\rm tot}= 10^6 $ and $ R_{\rm gc}=15~\rm pc$.  SGWB could be detected by CE for the mass ranges $40 M_\odot < m_1 \leq 55 M_\odot$ and 
    $14 M_\odot \lesssim m_2 \leq 25 M_\odot$ for the range  of small $r_i$ (gray area in the bottom-left plot of Fig.~\ref{fig:background}), while for the range of larger $r_i$ (orange area in the bottom-left plot of Fig.~\ref{fig:background}) detection is possible in the mass ranges $26 M_\odot < m_1 \leq 55 M_\odot$ and $9 M_\odot \lesssim m_2 \leq 20 M_\odot$.
    
    \item \label{lb:4} \textit{Case $4$: } We set $N^{\rm tot}= 5 \times 10^5 $ and $ R_{\rm gc}=15~\rm pc$. We find that the SGWB can be detected by CE  for the range of small $r_i$ (gray area in the bottom-right plot of Fig.~\ref{fig:background}) and within the mass ranges $33 M_\odot \lesssim m_1 \leq 55 M_\odot$ and $11 M_\odot \lesssim m_2 \leq 25 M_\odot$. Moreover, we observe that the SGWB can be detected via
CE for the range of larger $r_i$ (orange area in the right bottom plot of Fig.~\ref{fig:background}) within the mass ranges $20 M_\odot \lesssim m_1 \leq 55 M_\odot$ and
    $6 M_\odot \lesssim m_2 \leq 20 M_\odot$,  and is likely also  detectable with DECIGO, but barely
detectable with ET.

\end{enumerate}

\end{itemize}

%%%%
%\subsection{Global map of SGWB}\label{sec:global}
\subsection{Total contribution to the SGWB from weakly hyperbolic encounters}\label{sec:global}

\begin{figure}[t]
    \centering
\includegraphics[scale=.58]{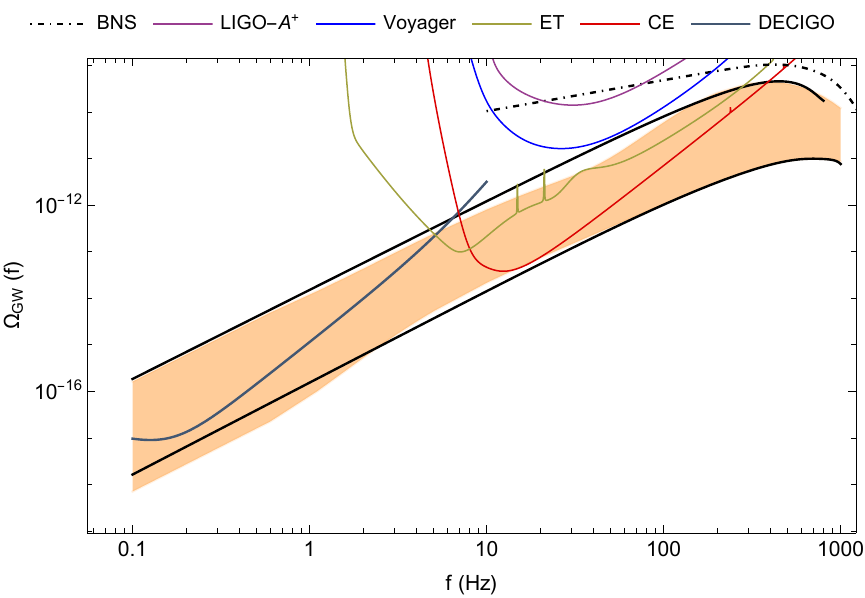}
    \caption{Total contribution to the SGWB from weakly hyperbolic encounters (orange area). In this plot, the virial velocity varies within $5 \rm{km/s} \leq v_i \leq 18 \rm{km/s}$. The black lines around the orange area show the fitted curves for this SGWB; the bottom line represents the pessimistic scenario, while the top line is the optimistic scenario (see Sect.~\ref{sec:global} for more details). Moreover, BNS, LIGO-$A^{+}$, and Voyager curves correspond to the energy density of GW background of the binary neutron star~\citep{bns}, and the sensitivity curves of the Advanced Laser Interferometer Gravitational-wave Observatory~\citep{ligoap}, and the Voyager \citep{LIGOScientific:2016wof,Sathyaprakash:2019rom} (see also \citep{ligodoc1}) detectors respectively for one year of observations.} 
    \label{fig:background5}
\end{figure}

\begin{table*}
\caption{The following table contains the fitting parameters for the optimistic and pessimistic scenarios for the SGWB from weakly hyperbolic encounters. See section~\ref{sec:global} for the definitions of each scenario.}
\centering
\begin{tabular}{ |p{1.5cm}|p{2.0cm}|p{2.5cm}|p{2.0cm}|p{2.5cm}|}
 \hline
 \multicolumn{5}{|c|}{Fitting parameters} \\
 \hline
 Parameters & Optimistic & Error (Optimistic) & Pessimistic & Error (Pessimistic)\\
 \hline
$a_0$ & $3.11 \times 10^{-22}$ & $4.54 \times  10^{-24}$ & $-2.87 \times 10^{-25}$ & $2.05 \times 10^{-28}$ \\
$a_1$ & $4.82 \times 10^{7}$ & $1.55 \times 10^{-52}$   & $-5.33 \times  10^{8}$ & $6.24 \times 10^{-61}$ \\
$a_2$ & $-10.8 \times 10^{4}$ & $9.26 \times 10^{-50}$   & $1.69 \times  10^{6}$ & $4.4 \times 10^{-58}$ \\
$a_3$ & $61.97$ & $6.36 \times 10^{-47}$   & $-3.5 \times 10^3$ & $3.34 \times 10^{-55}$\\
$a_4$ & $-$ & $-$   & $5.18$ & $2.63 \times 10^{-52}$\\
$a_5$ & $-$ & $-$ & $-0.004$ & $2.13 \times 10^{-49}$\\
$a_6$ & $-$ & $-$  & $1.4 \times 10^{-6}$ & $1.76 \times 10^{-46}$\\
$n$ & $0.92$ & $8.09 \times 10^{-45}$   & $0.98$ & $3.66 \times 10^{-52}$
\\
 \hline
\end{tabular}
\label{table_01}
\end{table*}

Here, we present the total contribution to the SGWB from all the previous cases. In particular, $n_{\rm co}(z)$ is the same as was defined in the previous section; we let $r_i$ vary in the whole range of $\lbrace r_{i\rm min}, r_{i\rm max} \rbrace= \lbrace 0.03\rm pc, 9.03\rm pc \rbrace$, and we set the same range for the eccentricity as above: $\lbrace e_{i \rm min}, e_{i\rm max} \rbrace= \lbrace e_{i \rm min}, e_{i \rm min}+ 4\times 10^{-9} \rbrace$. In section~\ref{sec:SpeCas}, we set $N^{\rm tot}$ and $R_{\rm gc}$ for each case, thus fixing the virial velocity~\eqref{eq:virialv}. However, for the total contribution of SGWB, we directly vary the virial velocity in the range $5 \rm{km/s} \leq v_i \leq 18 \rm{km/s}$. The resulting Figure~\ref{fig:background5} shows the SGWB for the mass ranges $17 M_\odot \lesssim m_1 \leq 55 M_\odot$ and $5 M_\odot \lesssim m_2 \leq 25 M_\odot$.

In order to provide a reference SGWB for hyperbolic encounters, which could be compared with other SGWB from other GW sources or other GW detectors, we provide a fitted analytical expression for the $\Omega_{\rm GW}(f)$ we find. Namely, from Figure~\ref{fig:background5}, we note that the plot follows a power law with a peak; we aim to fit this plot using the expression below:
%%%
\begin{equation}
    \Omega_{\rm GW}(f)=a_0 (1+\sum_{i=1}^{i_{\rm max}}a_i f^i)f^{n}.
\end{equation}
We perform two fits, one for the maxima $\Omega_{\rm GW}(f)$ of the orange area for each frequency in Figure~\ref{fig:background5}, which we refer to as the optimistic scenario, and one for the minima $\Omega_{\rm GW}(f)$ of the orange area for each frequency, which we refer to as the pessimistic scenario. In the optimistic scenario, we fit the SGWB using $i_{\rm max}=3$, while for the pessimistic scenario, we use up to $i_{\rm max}=6$. In  Table \ref{table_01}, we provide the fitting parameters for both scenarios.

%%%%%%%%%%%%
\section{Discussion} \label{sec:Disc}

In this work, we studied the SGWB from hyperbolic encounters inside bound compact clusters. Specifically, we investigate weakly hyperbolic encounters with eccentricities close to one, and compute the energy density of the SGWB. By considering some assumptions, given in Sect.~\ref{sec:StochBack}, we determine the energy density GW spectrum for a collection of hyperbolic encounters. 

In Sect.~\ref{sec:Results}, we present our findings and compare them with the sensitivity curves of GW detectors. First, to obtain insight into how certain parameters affect the spectra, we examine different scenarios. We consider two main cases: weakly hyperbolic encounters that occur close to the core of clusters and encounters starting from a radius close to the core out to a radius near the edge of a cluster. As shown in Sect.~\ref{sec:Results}, in both cases, the SGWB from these encounters can be detected mainly by the Cosmic Explorer GW detector. However, the detectability of these spectra depends on the virial velocity and the total mass of the system. By observing the different scenarios, we find the following trends:
\begin{itemize}
    \item As we accumulate more encounters, namely by considering more encounters from the core to the edge of the cluster, the chance of detectability increases.
    \item The chance of detectability increases by increasing the total mass $M$ of the system.
    \item Decreasing the virial velocity increases the chance of detectability.
\end{itemize}
Having these trends in mind, we provide a total contribution of SGWB from the weakly hyperbolic encounters, where we accumulate all cases and present an analytical expression for the most optimistic and pessimistic scenarios, respectively. We conclude that the SGWB from the weakly hyperbolic encounters appears to be detectable mainly by the Cosmic Explorer and DECIGO detectors, and a marginal detection may be possible with the Einstein Telescope.  

\begin{acknowledgements}
 MK, S. Mukherjee, and GLG have been supported by the fellowship Lumina Quaeruntur No. LQ100032102 of the Czech Academy of Sciences. S. Mukherjee is also thankful to the Inspire Faculty Grant DST/INSPIRE/04/2022/001332, DST, Government of India, for support. Moreover, S. Mukherjee and S. Mitra are grateful to DST, Government of India, for support under the Swarnajayanti Fellowship scheme. Finally, we would like to thank Sachiko Kuroyanagi for her advisement and remarks.
\end{acknowledgements}

\bibliographystyle{aa}
\bibliography{ref}

%%%%%%%%%%%%%%%%%

\appendix
\section{Hyperbolic motion}\label{sec:hypermotion}
For a hyperbolic motion with a given semi-major axis $a_c$ and semi-minor axis $b_c$ whose focus lies at $(h,0)$, we have the equation
%%%%
\begin{equation}\label{eq:hyperbolic}
    \frac{(x-h)^2}{{a_c}^2}-\frac{y^2}{{b_c}^2}=1\, .
\end{equation}
%%%%
By setting $h=a_c e$, where  $e$ is the eccentricity defined as $e= \sqrt{1+(b_c/a_c)^2}$, and using the transformation
\begin{equation}
    x= r \cos(\phi-\phi_0), \qquad y= r \sin( \phi-\phi_0),
\end{equation}
the hyperbola~\eqref{eq:hyperbolic} can be written in polar coordinates as
\begin{equation}
    r=\frac{p}{1+ e \cos(\phi-\phi_0)},
\end{equation}
where $p$ is the semilatus rectum given by $p= {b_c}^2/{a_c}= a_c (e^2-1)$.

By setting $a_c= a/\alpha$ and $b_c= b/\sqrt{\alpha}$, we arrive at
\begin{align}
    e&= \sqrt{1+\frac{b^2}{a^2} \alpha},\\
    p&= \frac{b^2}{a}= \frac{a}{\alpha} (e^2-1),\\
    r&= \frac{a (e^2-1)}{\alpha (1+ e \cos(\phi-\phi_0))}.
\end{align}
Thus, the trajectory given in Sect.~\ref{sec:orbitalparameter} is actually a part of the hyperbola~\eqref{eq:hyperbolic}, where $h= a_c e$, $a_c= a/\alpha$, and $b_c= b/\sqrt{\alpha}$.

\section{Derivation of the FT}\label{sec:ftderivation}

In section~\ref{sec:FT}, we apply the FT to Eqs.~\eqref{eq:q11}-\eqref{eq:q33} to derive the power radiation in the frequency domain, i.e., Eq.~\eqref{eq:powerf}. In this section, we present the details of this derivation.

As Eqs.~\eqref{eq:q11}-~\eqref{eq:q33} contain $\sinh {n\xi}$ and $\cosh{n \xi}$, to derive the FT of these equations we need to determine $\widehat{\sinh {n\xi}}$ and $\widehat{\cosh{n \xi}}$. Following~\citet{2020grobner}, one can get the following FTs:
\begin{align}
    \widehat{\sinh{n\xi}}&= \frac{n}{i \omega} e^{\nu \pi/2} e^{- n \pi i/2}\left[ K_{i \nu+n}(\nu e)+ e^{i \pi n} K_{i \nu-n}(\nu e) \right],\label{eq:ftsinhn}\\
    \widehat{\cosh{n\xi}}&= \frac{n}{i \omega} e^{\nu \pi/2} e^{- n \pi i/2}\left[ K_{i \nu+n}(\nu e)- e^{i \pi n} K_{i \nu-n}(\nu e) \right],\label{eq:ftcoshn}
\end{align}
where $\nu= \omega/\omega_0$ and
\begin{equation}
    K_\alpha (x)= \frac{1}{2} \boldsymbol{e}^{\alpha \pi i/2} \int_{-\infty}^{+\infty } \boldsymbol{e}^{\alpha \xi- i x \sinh \xi} d\xi
\end{equation}
is the modified Bessel function of the second kind. Consequently, for $n=1$ and $n=2$ Eqs.~\eqref{eq:ftsinhn} and~\eqref{eq:ftcoshn} reduce to
\begin{align}
    \widehat{\sinh \xi} &= -\frac{2 i}{\omega e } \boldsymbol{e}^{\nu \pi/2} K_{i \nu} ( \nu e),\label{eq:ftsinh}\\
    \widehat{\cosh \xi} &=\frac{2}{\omega} \boldsymbol{e}^{\nu \pi/2} K_{i \nu}^{\prime}(\nu e),\label{eq:ftcosh}\\
    \widehat{\sinh 2 \xi}&=-\frac{2}{ i \omega} \boldsymbol{e}^{\nu \pi/2}\left(\left(2-\frac{4}{e^2}\right) K_{i \nu}(\nu e)-\frac{4}{\nu e} K_{\nu e}^{\prime}(\nu e) \right) ,\label{eq:ftsinh2}\\
    \widehat{\cosh 2 \xi}&=\frac{8}{e \omega} \boldsymbol{e}^{\nu\pi/2}\left( K_{i \nu}^{\prime}(\nu e ) -\frac{1}{\nu e} K_{i \nu}(\nu e)\right),\label{eq:ftcosh2}
\end{align}
where 
\begin{equation}
    K^{\prime}_\alpha(x)= -\frac{1}{2}\left[ K_{\alpha-1}(x)+K_{\alpha+1}(x)\right].
\end{equation}
Now, by applying the transformation~\citep{arfken}
\begin{equation}
    K_\alpha(x)= \frac{\pi}{2} i^{\alpha+1} H_\alpha^{(1)}(ix),
\end{equation}
where $H_\alpha^{(1)}(i x)$ is the Hankel function, we arrive at
\begin{align}
    \widehat{\sinh \xi} &=  \frac{\pi}{\omega e}  H^{(1)}_{i \nu} (i \nu e) \label{eq:fthsinh},\\
    \widehat{\cosh \xi} &= - \frac{\pi}{ \omega}  H^{(1)\prime}_{i \nu} (i \nu e) ,\label{eq:fthcosh}\\
    \widehat{\sinh 2 \xi}&= \frac{\pi}{\omega} \left( \left(\frac{4}{e^2}-2\right) H^{(1)}_{i \nu} (i \nu e)+\frac{4 i}{\nu e} H^{(1)\prime}_{i \nu} (i \nu e)  \right),\label{eq:fthsinh2}\\
    \widehat{\cosh 2 \xi}&=- \frac{\pi}{\omega}\left(\frac{4 i}{\nu e^2} H^{(1)}_{i \nu} (i \nu e)+\frac{4}{e} H^{(1)\prime}_{i \nu} (i \nu e)\right),\label{eq:fthcosh2}
\end{align}
with 
\begin{equation}
    H_\alpha^{(1)\prime}(x)=\frac{1}{2}\left[H_{\alpha-1}^{(1)}(x)-H_{\alpha+1}^{(1)}(x) \right].
\end{equation}
We note that Eqs.~\eqref{eq:fthsinh}-~\eqref{eq:fthcosh2} are similar to the equations derived by~\citet{juan} up to a sign in $\sinh{n\xi}$. As the power radiation~\eqref{eq:powerf} contains the square of absolute values of these FTs, this sign difference does not affect the result. 
%%%%%%%

\end{document}